\documentclass[10pt,twocolumn,letterpaper]{article}

\usepackage{iccv}
\usepackage{times}
\usepackage{epsfig}
\usepackage{graphicx}
\usepackage{amsmath}
\usepackage{amssymb}
\usepackage{cite}

\usepackage[breaklinks=true,bookmarks=false]{hyperref}

\iccvfinalcopy 

\def\iccvPaperID{****} 

\ificcvfinal\fi

\begin{document}

\title{3D RegNet: Deep Learning Model for COVID-19 Diagnosis on Chest CT Image}

\author{Haibo Qi\\
Xidian University\\
Xian, China\\
{\tt\small hqi@stu.xidian.edu.cn}
\and
Yuhan Wang\\
Xidian University\\
Xian, China\\
{\tt\small yhwang\_0@stu.xidian.edu.cn}

\and
Xinyu Liu\\
Xidian University\\
Xian, China\\
{\tt\small liuxinyu@stu.xidian.edu.cn}
}

\maketitle
\ificcvfinal\thispagestyle{empty}\fi

\begin{abstract}
  In this paper, a 3D-RegNet-based neural network is proposed for diagnosing the physical condition of patients with coronavirus (Covid-19) infection. In the application of clinical medicine, lung CT images are utilized by practitioners to determine whether a patient is infected with coronavirus. However, there are some laybacks can be considered regarding to this diagnostic method, such as time-consuming and low accuracy. As a relatively large organ of human body, important spatial features would be lost if the lungs were diagnosed utilizing two dimensional slice image . Therefore, in this paper, a deep learning model with 3D image was designed. The 3D image as input data was comprised of two-dimensional pulmonary image sequence and from which relevant coronavirus infection 3D features were extracted and classified. The results show that the test set of the 3D model, the result: f1 score of 0.8379 and AUC value of 0.8807 have been achieved.
\end{abstract}

\section{Introduction}

The Coronavirus Disease 2019 SARS-CoV-2 (COVID-19) COVID-19 has spread all over the world, causing a major blow to the economies of all countries around the world\cite{sohrabi2020world,lai2020asymptomatic}. The early infection symptoms of the virus contain almost all the symptoms of common cold, and it can quickly aggravate the disease and be fatal. During the study, doctors found that in the early, middle and end stage, the condition of patient’s lung has very obvious change, and it can be seen from CT image.

In recent years, deep learning play a big role in the field of medicine , more and more CAD algorithms help doctors to diagnose diseases. Among them, deep convolution network is also widely used in the field of medical image, such as object detection on tumor\cite{zhang2019comparison}, semantic segmentation of lesion region\cite{rezaei2017conditional}, diagnosis of lymph node metastasis\cite{borrelli2021artificial}, etc\cite{wang2021predicting}. Various experiments have proved that AI can share the workload of doctors well. Pharyngeal swab detection is a common sampling method. But because of the strong infectivity of the virus, throat swab has a certain risk to the sampling doctors and people who are in close contact with it. Therefore, it has also increased the demand for AI assisted diagnosis. Based on this, we design an algorithm which can judge with lung CT image whether novel coronavirus pneumonia has infected in patients 

In traditional medical image classification tasks, for 2-dimensional image sequences, researchers often extract features from every 2D image, and then use RNN or other time series models to input the feature images in turn and finally obtain the diagnosis results\cite{kollias2020deep,kollias2018deep,kollias2020transparent}. Although this method is effective, not significant. Especially in the medical field, false negative usually causes greater risk than false positive. 

Radiologists usually regard the 2D diagram sequences of an organ as a three-dimensional diagram. In some time, they will observe the image of the whole organ from three angles. So for a pixel point, the local association of its six directions is equally important. In CNN+RNN model, due to the difference between the ideas of CNN and RNN, the features (from RNN) between pictures are different from the features (from CNN) extracted on the single picture. Therefore, we decided to use 3D image as a whole input and put it into 3D Convolutional Neural Network for training.

\begin{table*}[h]
    \begin{center}
        \setlength{\tabcolsep}{2mm}{
              \resizebox{1\hsize}{!}{
            \begin{tabular}{cccc|p{1.8cm}|p{1.8cm}|p{1.8cm}|p{1.8cm}|p{1.8cm}}
                \hline
                Model &
                num blocks &
                block width &
                group width &
                Optimizer &
                Loss Weight &
                Learning Rate &
                AUC &
                F1 score \\ \hline
    
                \cline{5-9}
                 & & & & Novo & 1.0 & 0.0001 & 0.8574 & 0.7996 \\ 
                \cline{5-9}
                  &  &   &   & SGD & 1.0 & 0.0001 & 0.8561 &  0.7938 \\
                \cline{5-9}
                  & {[}2,6,12,4{]} & {[}48,128,256,512{]} & {[}8,8,8,8{]} & Adam &  2.0 &  0.001 &  0.8398 &  0.8082 \\
                \cline{5-9}
                  3D-RegNet &  &   &   & Adam &  3.0 &  0.0001 &  0.8518 & 0.8104 \\
                \cline{5-9}
                  &  &   &   & Adam &  3.0 &  0.00001 &  0.8649 &  0.8158 \\ \cline{2-9}
                \cline{5-9}
                  & {[}1,3,7,4{]} & {[}48,128,256,512{]} & {[}8,8,8,8{]} & Adam &  3.0 &  0.0001 & 0.8302 &  0.7789 \\ \cline{2-9}
                \cline{5-9}
                  & {[}2,6,12,4{]} & {[}64,128,256,512{]} & {[}8,8,8,8{]} & Adam &  3.0 &  0.0001 &  0.7513 &  0.7345 \\ \hline
                \cline{5-9}
                Ensembled  & - & - & - & - &  - &  - &   0.8807 &  0.8397 \\ \hline
    
                \hline
            \end{tabular}
            }
            }
    \end{center}
    \caption{This table shows the parameter we chose to train the model. It can be seen that assemble models will get a higher F1 score than any single model. }\label{table1}
\end{table*}

\begin{figure}[t]
\begin{center}
\centerline{\includegraphics[height=8cm,width=8cm]{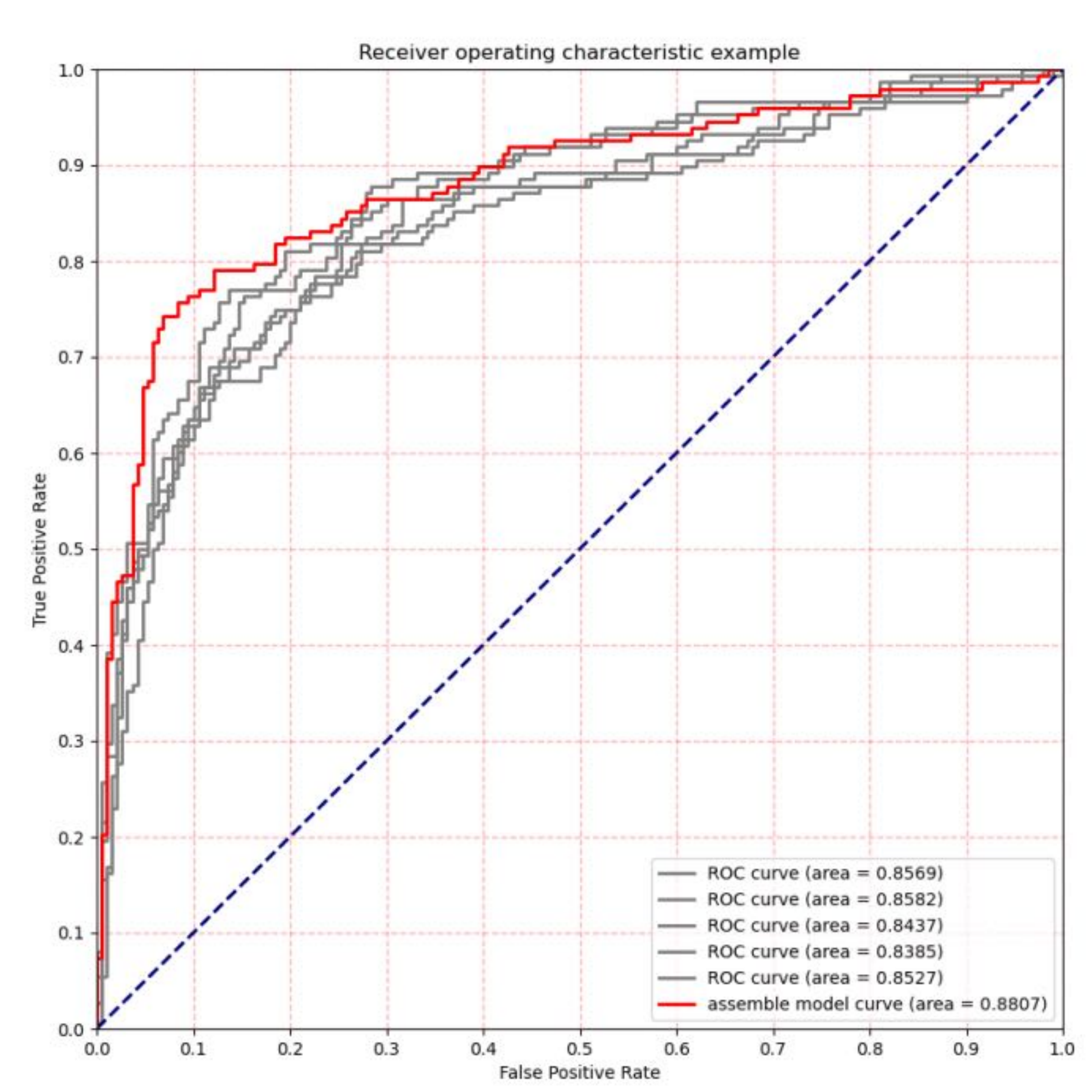}}
\end{center}
   \caption{This is the ROC curves of 5 models (in gray), and the fusion ROC curve of these models (in red). It can be seen that assemble models will get a higher AUC value than any single model.}
\label{fig:long}
\label{fig:onecol}
\end{figure}

\section{Related Work}

The COVID-19-CT-Database (COV19-CT-DB) consists of chest CT scans that are annotated for the existence of COVID-19\cite{kollias2021mia}. Data collection was conducted in the period from September 1, 2020 to March 31, 2021. Data were aggregated from many hospitals, containing anonymized human lung CT scans with signs of COVID-19 and without signs of COVID-19.

According to observation on the test set, we have found that the CT image sequence of a majority of patient is scanned from approximately the tibia all the way to the clavicle. To make sure the appropriate size of lung in the image, the top and bottom portions of the image are cropped to some extent, and the other four directions are also cropped to varying degrees.

Radiologists have reported a pattern of covid-19 infection with typical features including ground glass turbidity in the periphery of the lung, circular turbidity, and enlarged infiltrating vessels\cite{sohrabi2020world}. The presentation on CT images was darker areas of the lungs and distinct white streaks in the lungs of positive samples. For such images, we performed pre-processing methods on the data including increasing the contrast, to make a significant distinction between the infected and normal areas.
Before the data were put into the model, we utilized traditional data augmentation methods such as random cropping and horizontal flipping to prevent overfitting of the model. For the change of the size of the input image, a linear interpolation method was utilized to reduce or enlarge the image to the same size for training.
\section{Model}
The model we used to diagnosis COVID19 is Regnet\cite{radosavovic2020designing} with changes. We transform the original 2-dimensional convolution neural network into 3-dimensional convolution neural network. The architecture of our 3D Regnet is as shown in Figure 1.

3D regnet consists with stem (with the input of $w$,$h$,$d$ shape and output of $w$ $_{0}$ channels), body, and head (with the output of $n$ classes). As for the body part, it consists with few stages, and each stages are combined with identical standard residual bottleneck block\cite{he2016deep}. Each block consists of a 1×1×1 conv, a 3×3×3 group conv,  and a final 1×1×1 conv, where the 1×1×1 convs alter the channel width. BatchNorm and ReLU follow each conv. The block has 3 parameters: the width $w$ $_{i}$, bottleneck ratio $b$ $_{i}$, and group width $g$ $_{i}$. We have tried some combinations of the number of the blocks and the 3 parameters to use as our model. We are going to introduce the results in the Results part of this paper.For each patient, We put all the 2D images together to form a 3D image as the input of our model. In the last predicting stage, we use more than one models to predict the probability of each image, and take the average value as the final score. As a classification task, we need to have a threshold to split the probabilities into 2 classes. Therefore we draw the ROC curve with the predicting result of testing dataset, and take the threshold at the optimal cut-off point (find this point with Youden index) for classification.

Sequences of two-dimensional images of each patient are put together to form a three-dimensional image. Model fusion method was utilized in the pre-testing phase, where the images were put into multiple trained parametric models to predict the results and obtain the average values. For negative and positive classification, roc curves were plotted based on the prediction results of the test set samples and find the threshold value corresponding to the Jorden index point for classification.

\section{Experiments}

\subsection{Environment configuration}
We implemented our models on the PyTorch (version 1.7.1) deep-learning framework with a Linux 18.04 computer equipped with 4 NVIDIA TITAN V graphics processing units for training.

\subsection{Implementation Details}
3D images take very large space of memories, so that we used APEX which is a tool to provide the mixed precision calculation, to save computing space and increase the speed. We have trained few models under the different combinations of hyper parameter. The scores we get from these different models are shown in Table 1. 

We draw the ROC curve of the testing dataset, and take the threshold at the optimal cut-off point (find this point with Youden index) for classification. If the model gives the confidence which is higher/lower than threshold, we take it as covid/non-covid sample. Followed we calculate the F1 score under this threshold.

We used 4-7 of these models to do a model fusion. It proves that assemble model can obtain a better result than any of the single model. The AUC value of ROC curve and F1 score are improved by model fusion.

\section{ Results}

In this paper we have introduced a 3D RegNet model to diagnosis COVID-19. 3D model will be a new solution on medical 2D image series in future. Artificial Intelligence has been more and more widely used in medicine. Deep Learning model has improved the development of medical image field, no matter what kind of images, CT, MRI or Ultra Sound. A new deep learning algorithm equivalents to many experienced radiologists.

{\small
\bibliographystyle{ieee_fullname}
\bibliography{document.bib}
}

\end{document}